\documentclass[aps,prc,showpacs,showkeys,superscriptaddress,preprint]{revtex4}
\usepackage{amsfonts}
\usepackage{amsmath}
\usepackage{amssymb}
\usepackage{graphicx}
\usepackage{subfigure}
\begin{document}
\title{Continuum-continuum coupling and polarization potentials for weakly bound systesm}
\author{L.F. Canto}
\affiliation{Instituto de F\'{\i}sica, Universidade Federal do Rio de Janeiro, C.P.
68528, 21941-972 Rio de Janeiro, Brazil }
\author{J. Lubian}
\affiliation{Instituto de F\'{\i}sica, Universidade Federal Fluminense, Av. Litoranea
s/n, Gragoat\'{a}, Niter\'{o}i, R.J., 24210-340, Brazil}
\author{P. R. S. Gomes}
\affiliation{Instituto de F\'{\i}sica, Universidade Federal Fluminense, Av. Litoranea
s/n, Gragoat\'{a}, Niter\'{o}i, R.J., 24210-340, Brazil}
\author{M. S. Hussein}
\email{hussein@if.usp.br}
\affiliation{Instituto de F\'isica, Universidade de S\~{a}o Paulo,
  C. P. 66318, 05314-970 S\~{a}o Paulo, SP, Brazil}

\begin{abstract}
We investigate the influence of couplings among continuum states in collisions of 
weakly bound nuclei. For this purpose, we compare cross sections for complete fusion, 
breakup and elastic scattering evaluated by continuum discretized coupled channel (CDCC) calculations, 
including and not including these couplings. In our study, we discuss this influence 
in terms of the polarization potentials that reproduces the elastic wave function 
of the coupled channel method in single channel calculations. We find that the 
inclusion of couplings among continuum states renders the real part of the polarization 
potential more repulsive, whereas it leads to weaker absorption to the breakup channel. 
We show that the non-inclusion of continuum-continuum couplings in CDCC calculations 
may lead to qualitative and quantitative wrong conclusions. 
\end{abstract}

\maketitle

 After the elapse of almost two decades of extensive experimental
and theoretical effort, full understanding of the way the coupling to
the continuum influences the near-barrier fusion and other channels
in collisions of weakly bound nuclei is still lacking~\cite{CGD06,LiS05,KRA07}. 
Collisions of weakly bound projectiles can lead to different kinds of fusion.
The first is the usual complete fusion (CF), when the whole projectile is absorbed
by the target. The second type is incomplete fusion (ICF). This process corresponds to
the situation where the projectile breaks up into fragments along the collision and 
some fragments are absorbed while at least one is not. In addition, the breakup
process may end up as non-capture breakup (NCBU). In this case, none of the fragments is
absorbed. An ideal theoretical description of the collision should take into account
all these processes. In a rather detailed calculation within the Continuum Discretized Coupled Channel 
(CDCC) model ~\cite{SYK83,SYK86,AIK87}, Diaz-Torres and Thompson \cite{DiT02}, 
have managed to supply some very useful information about the aforementioned 
question. They found that the continuum coupling hinders the complete fusion 
cross section above and below the Coulomb barrier, with enhacement setting in 
only at deep sub-barrier energies. They further found that the inclusion of
the continuum-continuum couplings (CCC) was of paramount importance in
reaching the above conclusions about fusion. Their findings seem to
concur with experimental data~\cite{DHB99,DGH04}. Increased fusion can arise from
a lowering of the Coulomb barrier which results in a greater tunneling. 
On the other hand, decreased fusion can arise from
an increase in the height of the barrier which results in a smaller
tunneling. This latter effect would, in principle, be accompanied by
an increase in the quasi-elastic scattering at backward angles.

This ``common sense'' argument about the effects on the non-capture, quasi-elastic, processes
seem not be borne out by explicit calculation which takes into account the CCC 
effects~\cite{NuT99,LCA09}.  It is difficult to discern the physics behind all of the above. Clearly the 
inclusion or exclusion of the CCC dictates whether one is dealing with geniune 
breakup or merely a collection of inelastic channels. Further, there seems to be 
a need to invoke concepts such as irreversibility, and doorway that funnels
the flux to the continuum channel and hinders its return to the
entrance channel. Attempt to develop quantum transport theory for
these reactions, which incorporates these concepts at the outset, has recently
been made \cite{DHD08}. It becomes abundantly clear that more insight
into the working of the CCC and the resulting irreversibility and
doorway constriction is called for. This is an important issue which
goes beyond the realm of nuclear physics. In fact, as far back as
1961, Fano \cite{Fan61} elaborated a elastic + breakup theory for the
autoionization lines in atoms.

Our aim in this paper is to further elucidate the physics of the CCC in the 
context of reactions involving weakly bound nuclei at near-barrier energies. 
For this purpose, we investigate the effects of the CCC on the cross sections for CF, 
NCBU and elastic scattering. We find that the CCC lowers the CF and the NCBU
cross sections, but enhances the elastic cross section at large angles. The
calculated polarization potential clearly indicates a repulsive real
part and a reduced imaginary part.

In the CDCC method~\cite{SYK83,SYK86,AIK87}, the continuum states of the dissociated projectile
are approximated by a discrete set of wave-packets. In this way, the coupled-channel problem in
the continuum can be  handled analogously to the ones containing only bound channels. 
The scattering wave function is expanded in components with well defined values of total angular 
momentum  and its z-projection. The full Schr\"odinger equation is then projected on each intrinsic 
state and one gets a set of differential equations. The difference between the CDCC and a 
coupled-channel problem restricted to bound channels is that the configuration space of the 
former is much larger. The computational problem is then considerably more complex. The calculation 
is greatly simplified if it takes into account only the couplings among the bound channels and the
couplings between one bound channel and one channel in the continuum. 
In this way, continuum-continuum couplings are left out. This procedure was 
adopted in the first CDCC calculation of the CF cross section for the 
$^{11}$Be + $^{208}$Pb system, performed by Hagino {\it et al.}~\cite{HVD00}. 
The importance of CCC was investigated in a subsequent CDCC calculation for the same
system, performed by Diaz-Torres and Thompson~\cite{DiT02}. In this work, the CF cross sections evaluated with
and without CCC were compared. 
\begin{figure}[th]
\centering
\includegraphics[width=8 cm]{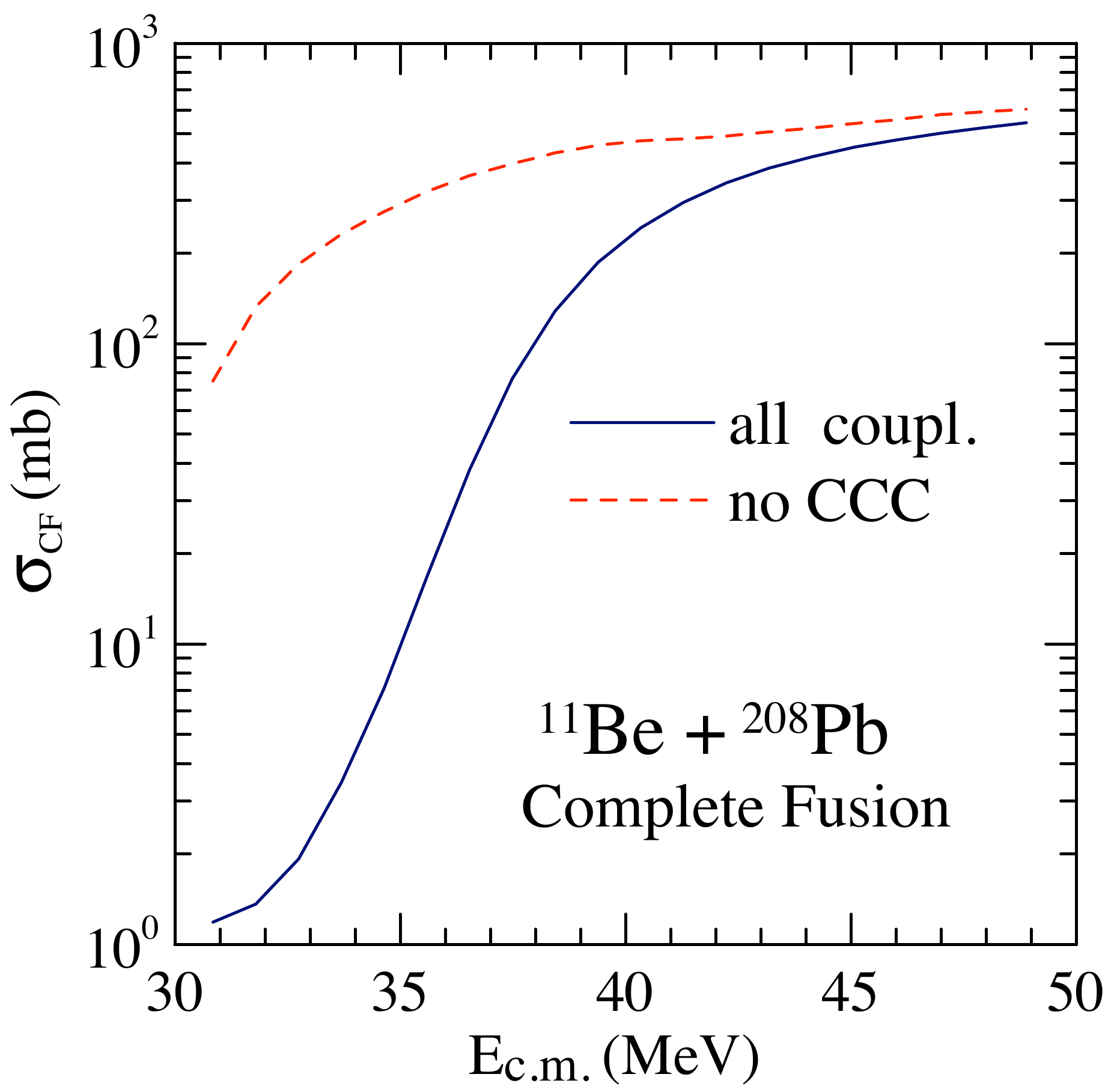}
\caption{(Color on line) CF cross sections in the $^{11}$Be + $^{208}$Pb collision. The solid 
and the dashed lines correspond respectively to CDCC calculations with and without CCC~\cite{DiT02}.}
\label{fusion}
\end{figure}
The results are shown in Fig.~\ref{fusion}. The comparison indicates that CCC leads to a drastic reduction 
of the CF cross section at near-  and sub-barrier energies. The reduction is of about two orders of magnitude.
\begin{figure}[th]
\centering
\includegraphics[width=7 cm]{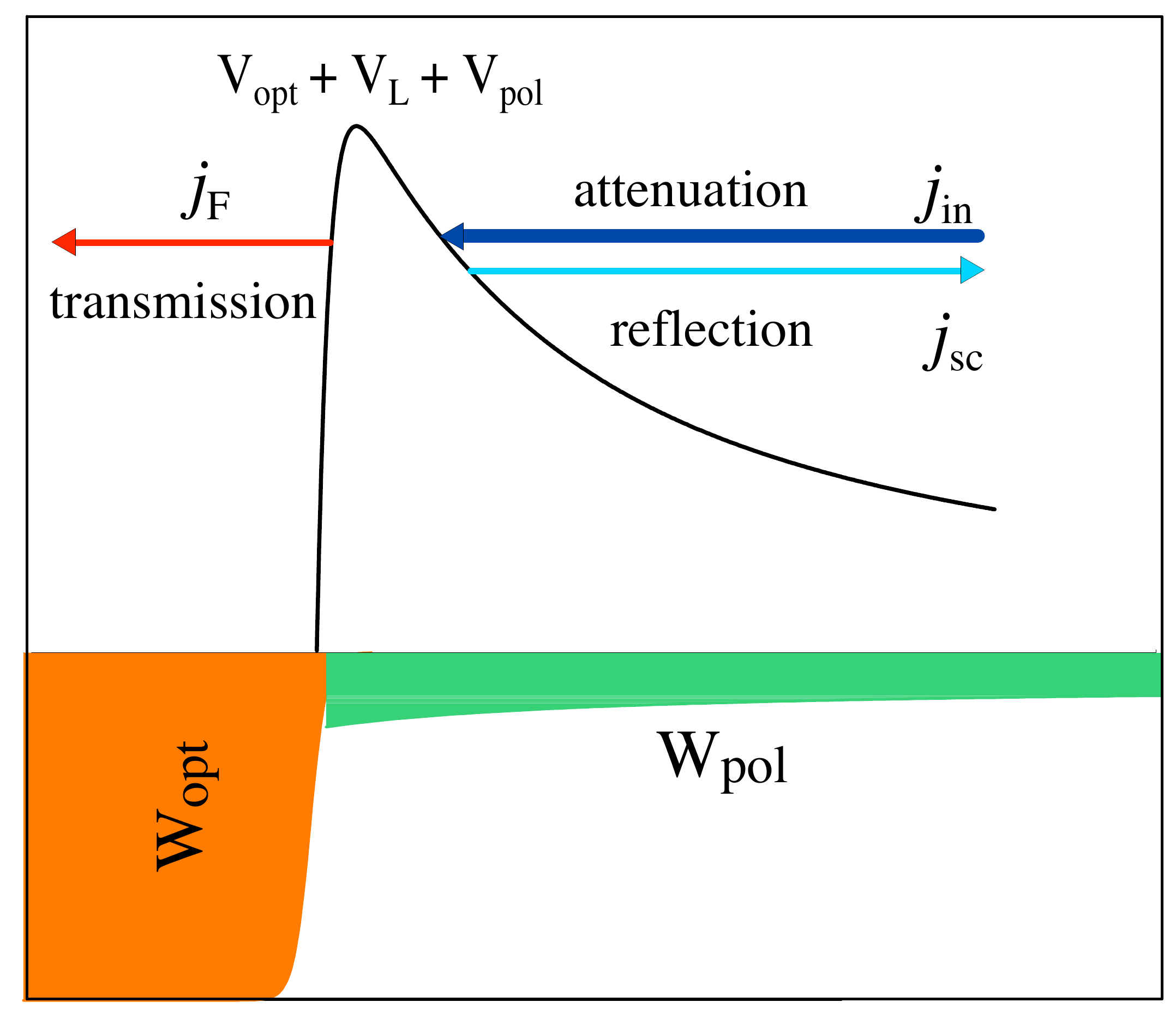}
\caption{(Color on line) Effects of the real and imaginary parts of the optical and the polarization potential on the
incident current.}
\label{schematic}
\end{figure}

Is there a simple and intuitive explanation for this result? To answer this question we use the language of
polarization potentials. The elastic wave functions obtained from a set of coupled channel equations
can always be obtained from a single-channel equation with an effective potential. This potential is
the sum of the optical potential and polarization potentials. The former represents the diagonal part of 
the interaction in channel space and an average influence of channel coupling. The latter contains
the detailed influence of the strongly coupled excited channels. In this way, the coupled channel problem
can be handled as a problem of potential scattering. Following the approach of Ref.~\cite{CCD02}, we 
use this picture and resort to the schematic representation of Fig.~\ref{schematic}. It shows the currents and the
potentials (real and imaginary parts) involved in the collision, for some particular partial wave. 
The fusion barrier is the sum of the real parts of the optical and polarization potentials, plus
the centrifugal term. As the incident current, $j_{\rm in}$, approaches the external turning
point, it is attenuated by the long range absorptive potential $W_{\rm pol}$. The lost flux populates
the channels that are responsible for this imaginary potential, that is, inelastic channels, transfer and
breakup. At large distances, $W_{\rm pol}$ is dominated by Coulomb breakup. The final destination 
of the fragments produced by the breakup process, namely NCBU, ICF is not relevant for our discussion. 
The situation would be different if all the fragments were absorbed sequentially, leading to CF.
Since the contribution of this process to the CF cross section is not supposed to be large, it is 
neglected here. When the attenuated incident current reaches the barrier, it splits into two parts. The reflected 
component, $j_{\rm sc}$, and the transmitted current, $j_{\rm CF}$. The reflected current is attenuated 
as it moves away from the barrier, until it is out of the reach of $W_{\rm pol}$. It is then responsible 
for the elastic scattering cross section. The transmitted current is fully absorbed by the short-range 
imaginary part of the optical potential inside the barrier, giving rise to CF. The probabilities of elastic 
scattering, $P_{\rm sc}$, and fusion $P_{\rm CF}$, at that partial wave then given by
are
\begin{equation}
P_{\rm sc}= \frac{j_{\rm sc}}{j_{\rm in}}
\qquad\qquad {\rm and} \qquad\qquad 
P_{\rm CF}= \frac{j_{\rm CF}}{j_{\rm in}}. \label{Psc-PF}
\end{equation}
The direct reaction probability, representing inelastic scattering + transfer + ICF + NCBU, 
is given by the current absorbed by $W_{\rm pol}$,
\begin{equation}
P_{\rm DR}= 1-\frac{j_{\rm sc} + j_{\rm CF}}{j_{\rm in}}.
 \label{Pqe}
\end{equation}

We can now speculate on the modifications of the polarization potential arising from the inclusion
of CCC in the CDCC calculations. The strong suppression observed in Fig.~\ref{fusion} implies that
the transmitted current is drastically reduced. In principle, it could be caused by three factors:
\begin{enumerate}
\item  the inclusion of CCC strengthens the absorptive imaginary potential $W_{\rm pol}$. In this case
the quasi-elastic cross section becomes larger, due to the increase of breakup.
\item  the inclusion of CCC makes the real part of the polarization potential more repulsive, so that the
incident current has to cross a higher barrier to produce fusion. If this case, $j_{\rm CF}$ is reduced
and $j_{\rm sc}$ increases. Therefore, the suppression CF should be followed by an enhancement of the elastic 
scattering cross section.
\item  a combination of possibilities 1 and 2. In this case, both the breakup and the elastic cross sections 
could be enhanced.
\end{enumerate}

\bigskip
\begin{figure}[th]
\centering
\includegraphics[width=9 cm]{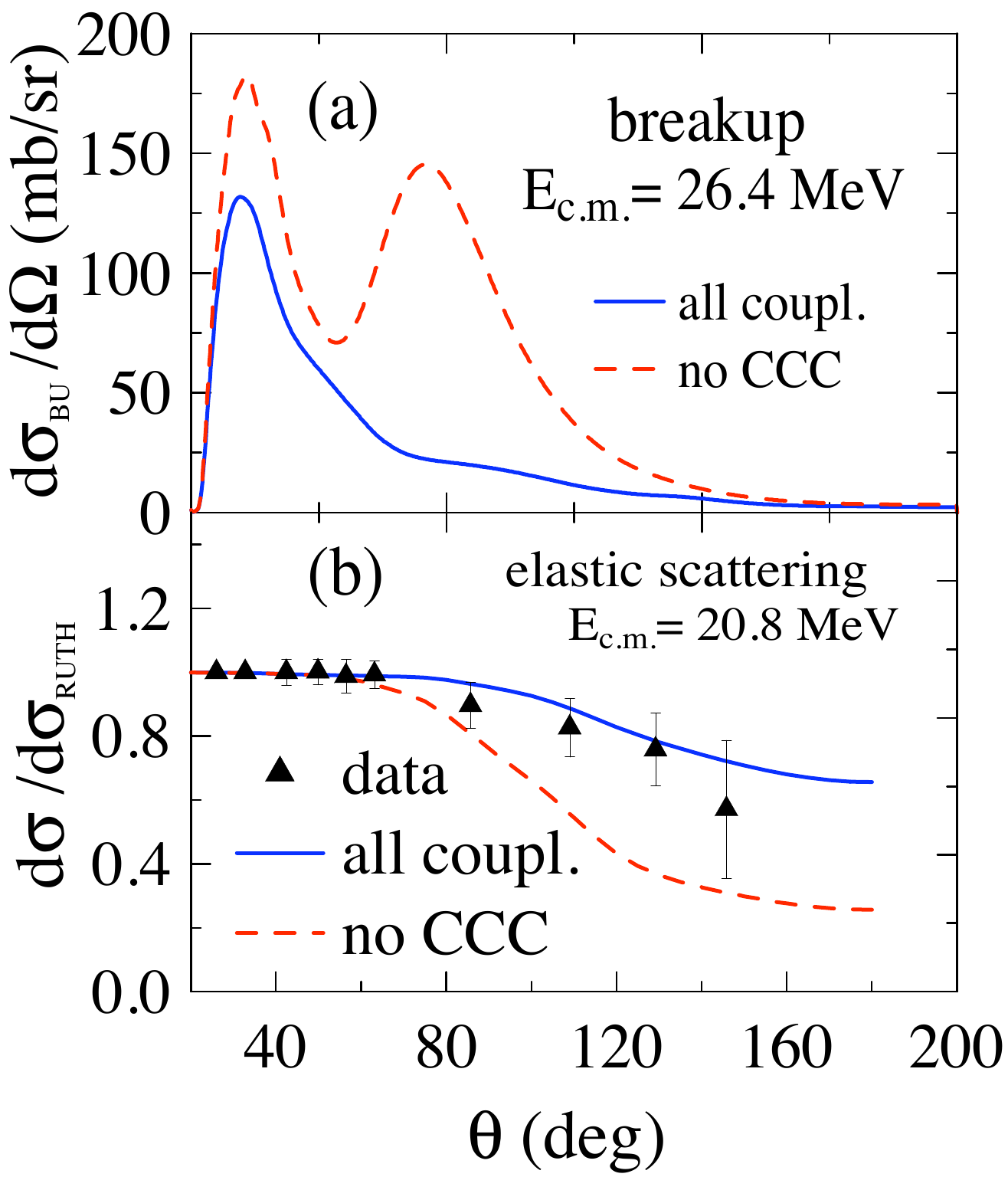}
\caption{(Color on line) Angular distributions for different processes in the $^8$B + $^{58}$Ni collision, calculated 
with (solid lines) and without CCC (dashed lines). In panels (a) and (b) we show, respectively, results for 
breakup~\cite{LuN07} and elastic scattering~\cite{LCA09}. For details, see the text.}
\label{breakup}
\end{figure}
To find out which of these possibilities is actually happening, one should check the cross section 
for other  channels. We first consider possibility 1. In this case, the reduction of $\sigma_{\rm CF}$
would arise from some kind of {\it  irreversibility} of the transition to the continuum. Of course, there is no 
irreversibility in quantum mechanics, as any transition can take place in two directions. However the 
elastic transition matrix is a superposition of a direct process with multistep processes of higher orders.  
To evaluate it, one should then perform sums over intermediate states. When CCC is taken into account, 
some of these sums become integrals. If these contributions have random phase, one could have destructive 
interference. Although this is a plausible hypothesis, there is no convincing arguments supporting it. 

To settle the matter, we first look at results of CDCC calculations for the NCBU
 cross section,  performed by Lubian and Nunes~\cite{LuN07}. In panel (a) of Fig.~\ref{breakup},
we show angular distributions of the center of mass of the $^8$B projectile in its breakup in the $^8$B + 
$^{58}$Ni collision. The solid and the dashed lines correspond respectively to results of calculations with
and without CCC. These curves are similar to the ones obtained in Ref.~\cite{NuT99} for a different collision energy. 
The difference is that we only show the curves involved in our discussion, leaving out other details of their 
calculations. Comparing calculations with and without CCC, we conclude that CCC leads to a substantial suppression
of the breakup cross section. Therefore, possibilities 1 and 3 can be ruled out. 

We are then inclined to believe that the reduction of the CF cross section arises from an increase of the
height of the potential barrier, when CCC is taken into account. This can be checked in an investigation
of the elastic cross section. Such CDCC investigation of the elastic angular distributions, which also
included the elastic breakup cross scetions, was in fact performed in the
past. In particular Sakuragi {\it et al.}~\cite{SYK83} went at length in calculating these
observables for the systems $^6$Li + $^{28}$Si and $^6$Li  + $^{40}$Ca at two laboratory
energies of $^6$Li: 99 MeV and 155 MeV. In the following we present our results for
the elastic angular distribution of the proton halo nucleus $^8$B on $^{56}$Ni target at
a much lower laboratory energy of 23.77 MeV, corresponding to $E_{\rm c.m.}$ = 20.8
MeV. The results are shown in panel (b) of Fig. 3. The calculations are the
same  as in Ref.~\cite{LCA09}.  In this case the excitations of the target were not included explicitly. 
We do not show results for other collision energies because they are qualitatively similar. The figure shows
clearly that the inclusion of CCC leads to a very strong enhancement of the cross section, as compared to
the results without CCC. This confirms that possibility 2 is the mechanism responsible for the suppression of
complete fusion. That is, the main effect of CCC is making the polarization potential repulsive.
\begin{figure}[th]
\centering
\includegraphics[width=8 cm]{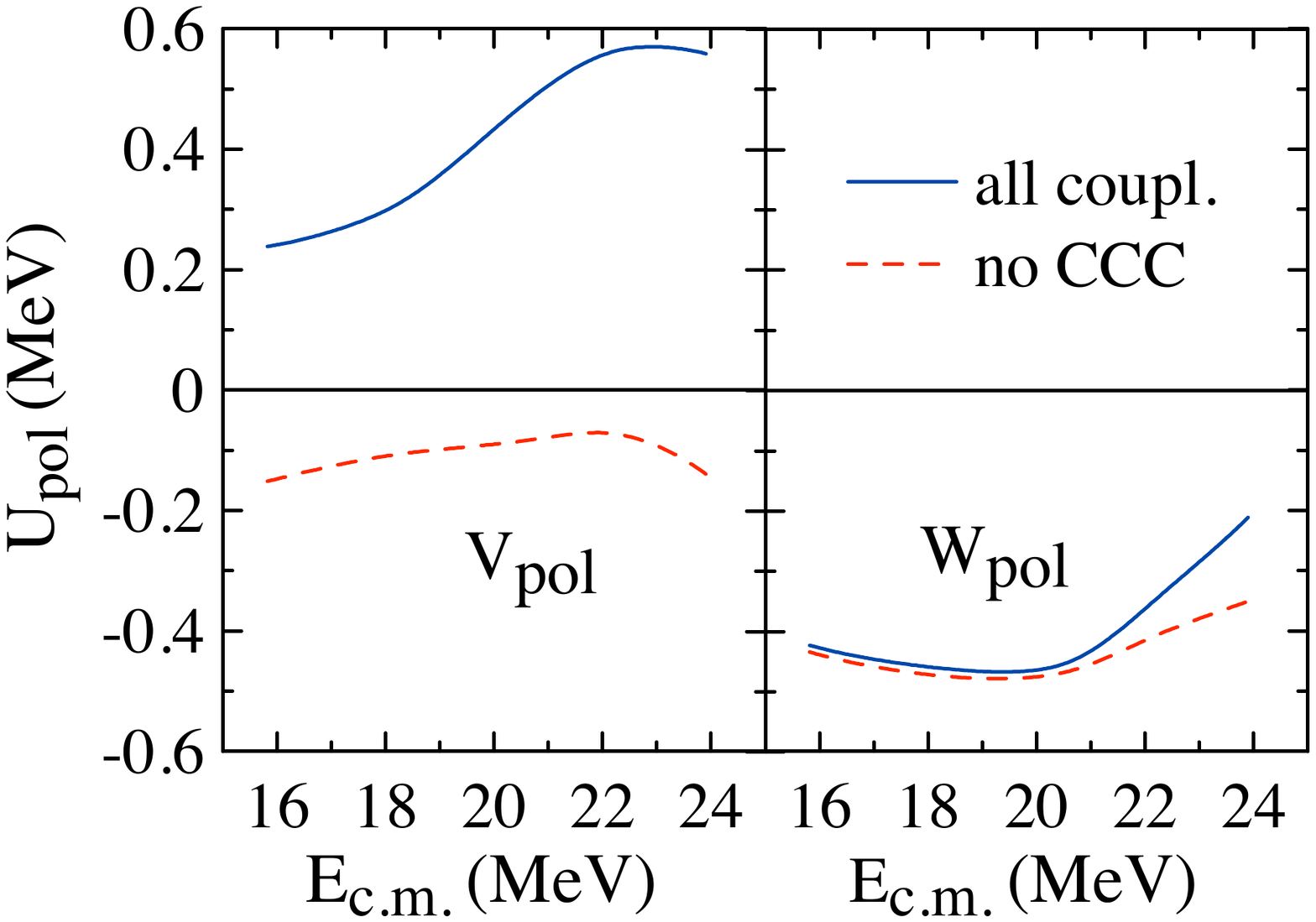}
\caption{(color on line)
Real (left panel) and imaginary (right panel) parts of the polarization potential for the $^8$B + $^{58}$Ni system.
Potentials based on CDCC calculations including and not including CCC are represented respectively by
solid and dashed lines.}
\label{Upol}
\end{figure}
\\

We now confront the conclusions of the previous section with the results of a direct analysis of the polarization
potential. The calculation of an exact polarization potential presents some serious difficulties. This potential 
depends strongly on the angular momentum and has poles. However, there are approximate angular momentum
independent polarization potentials which are free of poles and lead to reasonable predictions for cross sections.
We derive here the approximate polarization potential following the prescription of Thompson {\it et al.}~\cite{TNL89}. 
According to this prescription, the polarization potential is written as an average over angular momentum, involving
radial wave functions and S-matrix obtained from a coupled channel (in our case CDCC) calculation.  
We obtained polarization potentials based on CDCC calculations with and without CCC. The strengths of the real 
and imaginary parts of the polarization potential evaluated at the barrier radius are shown in 
Fig.~\ref{Upol}, for the $^8$B + $^{58}$Ni system. 
First, we note that the imaginary part of the polarization potential is alway negative, both 
in the calculations with and without CCC. This is not surprising since it represents the flux lost to the inelastic 
and breakup channels. The second relevant point is that the inclusion of CCC leads to a weaker imaginary potential,
reducing the absorption associated with direct reactions. This is consistent with the reduction of the NCBU cross section
found directly in our CDCC calculation with CCC. We now turn to the real part of the polarization potential. Again the effect of 
CCC on the polarization potential confirms our previous findings. In the absence of CCC, 
$V_{\rm pol}$ is attractive, reducing the barrier of the optical potential.  The inclusion of CCC modifies $V_{\rm pol}$ qualitatively. It becomes repulsive. In this way, the fusion barrier becomes higher and the CF cross section lower. The
higher barrier increases reflection and enhances the elastic elastic scattering cross section, as discussed in a
previous paragraph.

Our conclusions above are in complete qualitative agreement with those of Ref.~\cite{SYK83}, 
where the dynamic polarization potential was also fully investigated within CDCC taken into 
account the reorientation part of the the continuum-continuum coupling effects. These authors 
calculated the DPP for each orbital angular momentum and summed all the contribution 
(Eqs.~(7.4) and (7.5) of Ref.~\cite{SYK83}). They obtained for the total, $l$-summed DPP for 
$^6$Li + $^{28}$Si at $E_{{\rm Lab.}} = 99$ MeV a rather strong repulsive real part and a small 
imaginary part in the surface region. This is in agreement with our results described above.

Concluding, in this paper we investigated the role of the continuum-continuum
coupling in CDCC calculation of low-energy observables in heavy-ion reactions
involving weakly bound nuclei. We have found that this coupling reduces the
value of the non-elastic cross sections, which includes fusion, non-capture breakup, etc., 
in full agreement with previous works \cite{DiT02,NuT99}, while it increases
the purely elastic scattering cross section ratio to Rutherford
at back angles. We have traced this effect to the rather peculiar behaviour of
the breakup dynamic polarization potential which we found to have a repulsive
real part and a weaker absorptive part, when the CCC is included. Without the
CCC, the real part was found to be attractive with a stronger absorption. The
latter case is a common feature of coupling to inelastic channels, which leads us to
conclude that a discretized continuum can only be a loyal representative of a breakup
channel if the continuum-continuum coupling is fully accounted for in a CDCC
calculation.

\vskip 1cm
\noindent \textbf {Acknowledgements}  
%\medskip
\noindent We thank Alexis Diaz-Torres for supplying part of figure 1. This work was supported in part by the FAPERJ, CNPq, FAPESP and the PRONEX.

\end{document}